\documentclass[10pt,aps,twocolumn,prd,superscriptaddress,noshowpacs,nofootinbib,noshowkeys,floatfix]{revtex4-1}
\usepackage{graphicx}

\usepackage{multirow}
\usepackage{longtable}
\usepackage[normalem]{ulem}

\usepackage{booktabs}
\usepackage[hidelinks]{hyperref}
\usepackage{color}

\renewcommand\sout{\bgroup \color{blue} \ULdepth=-.5ex \ULset}
\hypersetup{
    colorlinks,
    linkcolor={blue},
    citecolor={blue},
    urlcolor={blue}
}
\usepackage{amssymb,amsmath,amsfonts}
\usepackage[utf8]{inputenc}
\usepackage{nicefrac}
\usepackage{upgreek}
\usepackage{braket}
\usepackage{slashed}

\begin{document}

\title{Fourier coefficients of the net baryon number density and their scaling properties near a phase transition}
\date{\today}

\author{G\'abor Andr\'as Alm\'asi}
\affiliation{Wigner Research Center for Physics, H-1121 Budapest, Hungary}
\author{Bengt Friman}
\affiliation{GSI Helmholtzzentrum f\"ur Schwerionenforschung GmbH,
D-64291 Darmstadt, Germany}
\author{Kenji Morita}
\affiliation{RIKEN Nishina Center, Wako 351-0198, Japan}
\affiliation{Institute of Theoretical Physics, University of Wroc{\l}aw, PL-50204, Wroc{\l}aw, Poland}
\author{Krzysztof Redlich}
\affiliation{Institute of Theoretical Physics, University of Wroc{\l}aw, PL-50204, Wroc{\l}aw, Poland}
\affiliation{Research Division and EMMI, GSI Helmholtzzentrum f\"ur Schwerionenforschung,
\\ 64291 Darmstadt, Germany}

\begin{abstract}
 We study the Fourier coefficients $b_k(T)$ of the net baryon
 number density in strongly interacting matter at finite temperature.
 We show that singularities in the complex chemical potential plane
 connected with phase transitions are reflected in
 the asymptotic behavior of the coefficients at large $k$. We derive the scaling
 properties of $b_k(T)$ near a second order phase transition in the
 $O(4)$ and $Z(2)$ universality classes. The impact of first order
 and crossover transitions is also examined. The scaling
 properties of $b_k(T)$ are linked to the QCD phase diagram in the
 temperature and complex chemical potential plane.
\end{abstract}

%\begin{keyword}
%\PACS
%\end{keyword}

\maketitle

\section{Introduction \label{sec:Introduction}}

Fourier decomposition is a useful technique for exploring
characteristic features of a system. In the study of hot and dense Quantum
Chromodynamics (QCD), Fourier decomposition of the grand partition
function for imaginary baryonic chemical potential has been used
to obtain the canonical partition function at fixed net baryon number.
One thus obtains thermodynamic quantities at nonzero (real) net
baryon densities, bypassing the sign problem in the calculations of
the grand partition function at real baryonic chemical
potentials~\cite{hagedorn85:_statis,Gibbs86,miller,hasenfratz92:_canononical,Barbour97,Forcrand2006,Alexandru2005,ejiri08:_canon_qcd,Boyda:2017dyo,Bornyakov:2017wzr}.

Recently, the Fourier decomposition of the net baryon number density
$\chi_1^B(T,\hat{\mu}_B) \equiv \partial(p/T^4)/\partial \hat{\mu}_B$
\begin{equation}
\label{eq:imagdens}
\textrm{Im}[
\chi_1^B(T,i\theta_B)] = \sum_{k=1}^{\infty} b_k(T) \sin(k\,\theta_B)
\end{equation}
at imaginary baryon chemical potential  $\theta_B=\text{Im}\,\hat{\mu}_B$
has been discussed as a tool which connects thermodynamic quantities at
imaginary and real chemical potentials. Here $\hat{\mu}_B=\mu_B/T$ is the reduced chemical potential
and $T$ the temperature.
In Ref.~\cite{Bornyakov:2017upg} it was shown that Eq.~\eqref{eq:imagdens} provides
a good description of the density obtained in lattice QCD simulations at imaginary $\mu_B$.
Given the density as a function of the imaginary chemical potential, one
may compute the Fourier coefficients $b_k$ using
\begin{equation}
 b_k = \frac{1}{\pi}\int_{0}^{2 \pi} \!\! d\theta_B \; \text{Im}\chi_1^B(T,i\theta_B)\sin(k\,\theta_B).\label{eq:bk}
\end{equation}
Note that Im$\chi_1^B(T,i\theta_B)$ is an odd, periodic function in
$\theta_B$ with the period $2\pi$.

Based on model calculations, it was shown in \cite{Kashiwa:2017swa}, that  the Fourier coefficients $b_k(T)$ provide a signature for the deconfinement transition. On the other hand, in Ref. \cite{Vovchenko:2017gkg}, an ansatz for the $k-$dependence of the Fourier coefficients
$b_k$ consistent with the lattice
data up to $b_4$ \cite{Vovchenko:2017xad} was employed to explore
the possible existence of a critical point at nonzero baryon density and to study fluctuations of the net baryon number.

\begin{figure}[!b]
 \centering
 \includegraphics[width=\columnwidth]{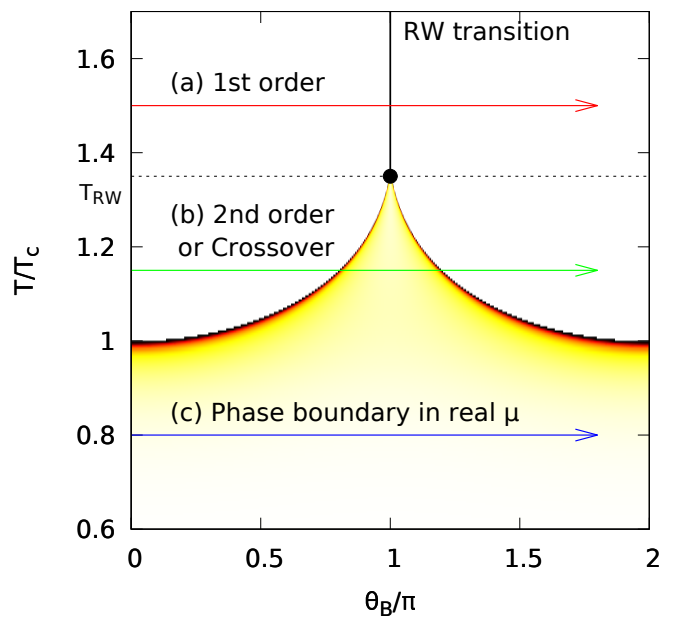}
 \caption{A schematic QCD phase diagram in the $T-\theta_B$ plane. The arrows indicate
 the integration paths in Eq.~\eqref{eq:bk} for high temperatures (a),
 for temperatures in the transition region (b) and for low temperature (c), respectively.}
 \label{fig:phasediagram_theta}
\end{figure}

The integration over the imaginary chemical potential in Eq. \ref{eq:bk} may cross a phase boundary, as indicated  in  Fig.~\ref{fig:phasediagram_theta}. The nature of the phase boundary depends on the temperature and on the pion mass. For instance,  at high temperatures,  QCD exhibits a first order phase transition at $\theta_B=\pi$, the Roberge-Weiss (RW) transition. The latter is a consequence of the $Z(3)$ center symmetry, which is related to the confinement of quarks~\cite{roberge86:_gauge_qcd}. The discontinuity of the baryon density at the RW transition is reflected in a power-law dependence of the Fourier coefficients, $b_k \sim 1/k$, for large $k$.
Moreover, even if the integration path does not cross a phase boundary,  as in case (c), or the phase boundary in case (b) is a crossover transition, in which case the density is analytic along the integration path, there are singularities in the complex chemical potential plane, which do affect the behavior of the Fourier coefficients.

The goal of this paper is to derive model independent results for the effect of critical singularities on the asymptotics of the Fourier coefficients\footnote{We note that although all singularities in the complex $\mu_B$ plane contribute to the Fourier coefficients, their asymptotic behavior is determined only by the singularity closest to the imaginary axis. Thus, the contribution to the asymptotics from the critical singularities considered in this letter, may be ``screened'' by other non-analyticities, located closer to the imaginary $\mu_B$ axis. Moreover, for singularities on the imaginary axis, the strongest singularity determines the asymptotics of the Fourier coefficients. Thus, again the contribution of a particular critical singularity may be eclipsed by other, stronger ones.}. First we study the Fourier expansion of the net baryon density in the  Landau theory of the phase transitions, in order to illustrate in a transparent framework how critical singularities in the complex chemical potential plane are reflected in the Fourier coefficients. We then apply the scaling theory of phase transitions to derive the properties of the Fourier coefficients near a second-order phase transition.
We also consider the influence of a first-order phase transition and
a crossover on the asymptotic behavior of the $b_k(T)$. An explicit calculation of the Fourier coefficients in a QCD-inspired effective model were presented in Ref.~\cite{Almasi:2018lok}.

In addition to the singularities associated with phase transitions, the thermodynamic potentials have thermal singularities~\cite{Skokov:2010uc}. In particular, the thermal branch points of the net baryon density are generated by the poles of the Fermi-Dirac function at complex values of the baryon chemical potential. For baryonic degrees of freedom of mass $m$, these are located at $\hat{\mu}_B=\pm \hat{m}\pm  i \pi$, where $\hat{m}=m/T$. The influence of the thermal singularities on the Fourier coefficients is also briefly discussed.

We present our results in the context of the expected criticality of QCD matter in the temperature and imaginary chemical potential plane (see Fig.~\ref{fig:phasediagram_theta}). In particular, we show that at the chiral $O(4)$ and the RW $Z(2)$ critical points, the asymptotic Fourier coefficients exhibit a power-law behavior, which depends on the critical exponents $\alpha$ and $\delta$, respectively. Furthermore, in the temperature range between the $O(4)$ critical point and the RW endpoint, an oscillation is superimposed on the power-law dependence. The frequency of the oscillation is determined by the value of the imaginary chemical potential at the critical point.

For temperatures below the $\mu_B\!=\!0$ chiral critical point, the $O(4)$ singularity is, in the chiral limit, located on the real $\mu_B$ axis. This implies that the contribution of the critical singularity to large-order Fourier coefficients is exponentially suppressed. Therefore the critical behavior at real $\mu_B$ does not have a conspicuous impact on the asymptotic form of the Fourier coefficients $b_k$. In particular, the existence of a $Z(2)$ critical point at large real $\mu_B$ is not perceptible in the high-order Fourier components.

Moreover, for physical quark masses, the $O(4)$ chiral transition is of the crossover type. In this case the critical singularities are located in the complex $\hat{\mu}_B$ plane and the $k$ dependence of the Fourier coefficients $b_k(T)$ corresponds to damped oscillations superimposed on a power-law dependence. The damping rate is determined by the real and the oscillation frequency by the imaginary part of the baryon chemical potential at the crossover branch point singularity. At temperatures between the chiral transition at $\mu_B=0$ and the RW critical point, the imaginary part dominates and the oscillations are underdamped. Conversely, at lower temperatures the real part of $\mu_B$ dominates and the oscillations are overdamped.

This paper is organized as follows. In section \ref{sec:landau} we discuss the analytic structure of the density in Landau theory. We then derive the asymptotic behavior of $b_k(T)$ in section \ref{sec:bk}. Finally, section \ref{sec:summary} is devoted to a summary.

\section{Critical singularities in the complex $\mu_B$ plane}
\label{sec:landau}

In the chiral limit of QCD, one expects to find a second order chiral
critical line, belonging to the $O(4)$ universality class,
at a critical temperature $T_c\simeq 140$ MeV and $\mu_B=0$~\cite{Ding:2018auz} and by continuity at small values of the baryon chemical potential.
This line continues to larger $\mu_B$ and ends in a chiral tricritical point at $(T_{TCP},\mu_{TCP})$, assuming that such a point exists. The second-order critical points and the tricritical point correspond to singularities on the real $\mu_B$ axis.

Given the negative curvature of the phase boundary at small $\mu_B$ and nonzero quark masses~\cite{Kaczmarek:2011zz,Endrodi:2011gv}, it is generally assumed that $T_{TCP}<T_c$. Beyond the tricritical point, i.e., at still larger values of the baryon chemical potential, the chiral transition would be first order.

On the other hand, at temperatures $T>T_c$, one expects a
chiral critical point for purely imaginary values of the chemical
potential, as shown in Fig.~\ref{fig:phasediagram_theta}. With increasing
temperature, this transition occurs at larger values of $\theta_B$ and
eventually approaches ${\theta_B=\pi}$, at or close to the temperature
of the Roberge-Weiss endpoint.

For nonzero quark masses, chiral symmetry is explicitly broken and the chiral critical line
is replaced by a line of crossover transitions. The corresponding branch points in the complex chemical potential plane are shifted
away from the real or imaginary axes to\footnote{The singularities come in complex conjugate pairs, because the thermodynamic functions are real for real values of the chemical potential. Moreover, for each singularity at $\mu_{br}$, there is one at $-\mu_{br}$, owing to charge conjugation symmetry.} $\mu_{br} = \pm\, \mu_c \pm i\, T\, \theta_c$ (see Fig.~\ref{fig:br}).  The conjectured first-order chiral  transition at large $\mu_B$ then ends in a
critical point  $(T_\text{CP},\mu_{CP})$, located on the real $\mu_B$ axis. This point is also referred to as the chiral critical endpoint~\cite{asakawa89:_chiral_restor_at_finit_densit_and_temper,Stephanov:1998dy}. 

\subsection{Landau theory}
\label{sec:landau-theory}

The Landau theory of  phase transitions offers a transparent framework for exploring the qualitative features of the singularity structures discussed above.
The Landau free energy is given by
\begin{equation}
 \Omega[\sigma] = \frac{1}{2}\,a\, \sigma^2 +\frac{1}{4}\,b\, \sigma^4 - h\,\sigma,\label{eq:potential}
\end{equation}
where $\sigma$ is the order parameter, $a$ and $b$ are real functions of the thermodynamic variables (here $T$ and $\mu_B$), while $h$ is a symmetry breaking external field\footnote{It is convenient to work in a dimensionless representation of \eqref{eq:potential}. This can be obtained,  e.g., by rescaling each quantity by appropriate powers of the critical temperature $T_c$.}.
The equilibrium value of the order parameter is determined by solving the gap equation,
\begin{equation}
 \frac{\partial \Omega}{\partial \sigma}=a\,\sigma+b\,\sigma^3-h = 0.
\end{equation}
By taking a further derivative with respect to $\sigma$, we obtain the order-parameter
susceptibility
\begin{equation}
 \chi_\sigma = (a+3\,b\,\sigma^2)^{-1},\label{eq:sus}
\end{equation}
where the order parameter takes on the value that minimizes $\Omega$.

The solutions of the gap equation read
\begin{align}
 \sigma_1 &= \frac{f(a,b,h)}{\sqrt[3]{2}\, 3^{2/3}\, b} -\frac{\sqrt[3]{\frac{2}{3}}\, a}{f(a,b,h)},\label{eq:sol1} \\
 \sigma_2 &= \frac{\left(1+i\sqrt{3}\right) a}{2^{2/3}\, \sqrt[3]{3}\,
 f(a,b,h)}-\frac{\left(1-i \sqrt{3}\right)f(a,b,h)}{2\,\sqrt[3]{2}\, 3^{2/3}\, b}, \\
 \sigma_3 &= (\sigma_2)^*.\label{eq:sol3}
\end{align}
where $f(a,b,h) = \sqrt[3]{9\, h\, b^2+\sqrt{3}\, \sqrt{27\, h^2\, b^4+4\, a^3\, b^3}}$.
The first solution is real, while the other two are
real or complex, for real $a$, $b$ and $h$.
In the limit $h\to 0$, where the symmetry under $\sigma\to -\sigma$ is restored, the
solutions reduce to $\sigma=0$ and $\sigma= \pm i\sqrt{a/b}$. The first one is a
minimum of the effective potential for $a>0$, while the other solutions corresponds to two
degenerate minima for $a<0$. Thus, for $a<0$ the expectation value of the order
parameter is nonzero and the symmetry is spontaneously broken.

While for $h=0$ the critical point is located at $\sigma=0$ and
$a=0$, for nonzero $h$ the singularities are shifted to complex values
of $\sigma$ and $a$. These singularities are branch points  where the
susceptibility \eqref{eq:sus} diverges for a solution of the gap
equation. For the three solutions
\eqref{eq:sol1}--\eqref{eq:sol3}, these branch points are found at~\cite{friman12:_phase_trans_at_finit_densit}
\begin{align}
 a_{\text{br1}} &= -3 \left( \frac{bh^2}{4} \right)^{1/3}, \\
 a_{\text{br2,3}} &= 3 e^{\pm i\pi/3} \left( \frac{bh^2}{4} \right)^{1/3}.\label{eq:a_br2}
\end{align}
Thus,  the location of the branch points $a_{\text{br}}$ scale with $h^{2/3}$.
We note  that the first one, $a_{\text{br1}}$,  is located on an
unphysical Riemann sheet in the complex $a$-plane, while
the two complex singularities \eqref{eq:a_br2} are responsible for the
crossover transition at real $a$ and nonzero $h$.

\begin{figure}[!t]
 \centering
 \includegraphics[width=\columnwidth]{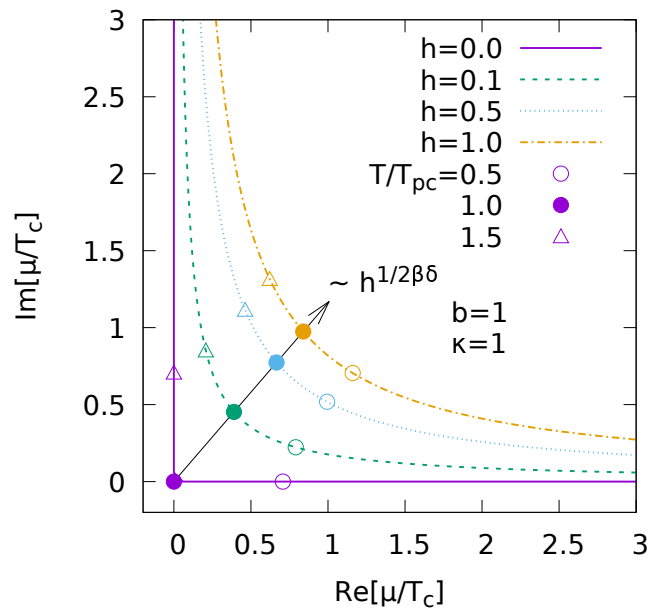}
 \caption{Movement of the branch point singularity with varying temperature in the first quadrant
 of the complex $\mu_B$ plane for $h=0$, 0.1, 0.5 and 1, computed using (\ref{eq:mu_br}). The rest of
 parameters is set to $b=1$ and $\kappa=1$. The symbols stand for
 the location of the branch point at $T/T_{\text{pc}}=0.5$, 1, and 1.5
 for each values of $h$.}
 \label{fig:br}
\end{figure}

To map the singularities onto the complex chemical potential plane,
we use the parameterization $a \equiv t^\prime =  t+ \kappa (\mu_B/T_c)^2$,  where $\kappa$ represents
the curvature of the critical line, $t=T/T_c - 1$ is the reduced temperature and $T_c$ the critical temperature at $\mu_B=0$. Then, the location of the four branch points is given by
\begin{equation}
 \mu_{\text{br}} = \pm\frac{T_c}{\sqrt{\kappa}}\sqrt{-t +
  \frac{3}{2}(1\pm\sqrt{3}i) \left( \frac{bh^2}{4} \right)^{\frac{1}{3}}}.\label{eq:mu_br}
\end{equation}
We note that for $t=0$, $\mu_{\text{br}}$ scales with $h^{1/3}$. As
observed above, there are four branch points in the
complex $\mu_B$ plane. In the chiral $(h\rightarrow 0)$ limit, the  branch points approach the real $\mu_B$ axis for $t < 0$ (below $T_c$) and the imaginary axis for $t > 0$ (above $T_c$), as indicated in
Fig.~\ref{fig:br}. In the chiral limit of QCD, the corresponding branch points
make up the $O(4)$ critical
line~\cite{stephanov06:_qcd_critic_point_and_compl,Skokov:2010uc,friman12:_phase_trans_at_finit_densit}.

For the crossover transition, one can define a pseudo\-critical
temperature at the maximum of the order parameter susceptibility
\eqref{eq:sus}. The solution of $\partial \chi_\sigma/\partial a=0$ is, for
the first solution of the gap equation \eqref{eq:sol1},
$a=3(bh^2/16)^{1/3}$, while the corresponding pseudocritical temperature at $\mu_B=0$ is given by
\begin{equation}\label{eq:Tpc}
 T_\text{pc} = T_c \left[ 1+ 3 \left( \frac{bh^2}{16} \right)^{1/3} \right]
\end{equation}
Using \eqref{eq:Tpc} in \eqref{eq:mu_br}, one finds the location of
the singularities at the pseudocritical temperature\footnote{As indicated in Fig.~\ref{fig:br}, $\mu_{\text{br}}(T=T_{\text{pc}})$ scales with $h^{1/2\beta\delta}$, which with mean-field exponents, $\beta=1/2$ and $\delta=3$, yields $h^{1/3}$.}
\begin{align}
 \mu_{\text{br}}(T=T_{\text{pc}}) &=
  \pm T_c\sqrt{\frac{3\,b\,(1-\sqrt[3]{2} \pm \sqrt{3} i )}{ \sqrt[3]{32}\,\kappa}}\,h^{1/3}\label{eq:mubr_tpc}\\
  &\simeq \, T_c\,\sqrt{\frac{b}{\kappa}}\,(0.839\pm 0.975\,i)\,h^{1/3}\nonumber
\end{align}
The trajectories of the crossover branch point in the first quadrant of the complex $\mu_B$-plane
are shown in Fig.~\ref{fig:br} for $h=0.1$, 0.5 and 1. The circles and triangles indicate the
locations of the branch point for fixed values of $T/T_{\text{pc}}$.

By adding a $\sigma^6$ term to the Landau free energy \eqref{eq:potential}, one can
explore the singularities associated with a tricritical point for $h=0$ and a critical point for $h\neq
0$. In the latter case, the two complex conjugate branch points of the crossover transition
merge on the real axis~\cite{friman12:_phase_trans_at_finit_densit} at $T=T_{\text{CP}}$. The possible existence of such a branch point,
which in QCD would correspond to a critical point belonging to the $Z(2)$ universality
class, is under intensive scrutiny in theoretical and experimental studies of nucleus-nucleus collisions~\cite{asakawa89:_chiral_restor_at_finit_densit_and_temper,stephanov06:_qcd_critic_point_and_compl,luo16:_explor_qcd}.

In lattice calculations it is found that the RW endpoint
is a triple point for large and small quark masses
\cite{forcrand:_const_qcd}. In the latter case, the chiral transition at imaginary $\mu_B$
is first order close to the RW endpoint, and may remain first order up to $\mu_B=0$ and beyond,
i.e., also for real values of $\mu_B$. Recent lattice results
suggest that for physical quark masses the RW endpoint is a second-order critical point
belonging to $Z(2)$ universality class~\cite{Bonati:2016pwz}. Thus, depending on the value of the quark mass,  the chiral
transition at imaginary $\mu_B$, i.e., in the temperature range $T_\text{pc}\leq T\leq T_\text{RW}$,
can be first order, of the crossover type or partly first order, partly crossover with a critical point.
In the following we explore the characteristics of the Fourier coefficients for these possibilities.

\subsection{Scaling theory}

This discussion can be generalized to properly account for critical fluctuations using the
scaling theory of phase transitions.
The singular part of the free energy is expressed in terms of the universal scaling function
$f_f(z)$
\begin{equation}
 f^{\text{sing}}(t,h) \sim h^{1+1/\delta}f_f(z)\label{eq:singular_f},
\end{equation}
where
$z\equiv t^\prime/h^{1/\beta\delta}$ is the scaling variable.

The scaling function exhibits universal branch points in the complex $z$ plane at $z=z_{\text{br}}$ and $z=z^*_{\text{br}}$.
For the corresponding singularities in the complex $\mu_B$ plane, we find, using the parametrization for $t^\prime$ introduced above,
\begin{equation}
 \mu_{\text{br}} = \pm T_c \left[ \frac{1}{\kappa}(z_{\text{br}}-z_0) \right]^{1/2}h^{1/2\beta\delta}\label{eq:mubr_scaling},
\end{equation}
where $z_0=(T/T_c-1)/h^{1/\beta\,\delta}$. The two remaining singular points are obtained by replacing $z_{\text{br}}$ by $z^*_{\text{br}}$
in (\ref{eq:mubr_scaling}). The behavior of the singular parts of the free energy and the baryon
density close to the branch point is obtained by analyzing \eqref{eq:singular_f}.

We note that singularities of the thermodynamic functions influence the
asymptotic behavior of the corresponding Fourier series obtained at imaginary values of the
chemical potential~\cite{ZinnJustin:2007zz,Tolstov:1976}. By appropriately deforming the integration contour in the
complex $\mu_B$ plane, one can isolate the contribution of each individual
singularity to the Fourier coefficients. In the following,
we determine the asymptotic behavior of the Fourier series stemming
from the singularities associated with phase transitions by using the
universal properties of the free energy.

\section{Asymptotics of Fourier coefficients}
\label{sec:bk}

For the discussion in this section, the Riemann-Lebesgue (RL) lemma~\cite{Tolstov:1976}
is of central importance. Applied to Eq.~\eqref{eq:bk}, it states that the Fourier coefficients
$b_k \rightarrow 0$ for  $k\rightarrow \infty$,
provided the integral of the density $\chi_1^B$ over the same interval exists.
In the appendix, the Riemann-Lebesgue lemma is proven for the more restricted case 
of a differentiable function.

\subsection{First order transition at imaginary $\mu_B$}
Let us first consider the case when the density has a discontinuity at
$\theta_B=\pi$. This corresponds to the RW phase transition in QCD at
high temperature, as indicated by case (a) in
Fig.~\ref{fig:phasediagram_theta}. At temperatures below
$T_{\text{RW}}$ the density vanishes at $\theta_B=\pi$,  while above $T_{\text{RW}}$ it is non-zero,
$\text{Im}\chi_1^B(T>T_{\text{RW}},i\theta_B=i\pi) \neq 0$.
Since $\chi_1^B(T,i\theta_B)$ is odd in $\theta_B$ and periodic under $\theta_B\to \theta_B+2\pi$, this implies that the density must be
discontinuous in this point. Using the periodicity and the symmetry of the integrand in \eqref{eq:bk}, one finds after integration by parts,
 \begin{align}\label{eq:RW-1st}
  b_k&= \frac{2}{\pi}\int_0^\pi d\theta_B \big[\mathrm{Im}\chi_1^B(T,i\theta_B)\big]\sin(k\,\theta_B) \nonumber \\
  &= 2\frac{ (-1)^{k-1}}{\pi k} \chi_1^B(T,i\theta_B=i\pi)  \\
  &\,  +\frac{2}{\pi k}\left(\int_0^\pi d\theta_B
  \big[\mathrm{Re}\chi_2^B(T,i\theta_B)\big]\cos(k\,\theta_B)\right).\nonumber
 \end{align}
 The expression in parentheses in the last line vanishes for
 $k\rightarrow \infty$ due to the RL lemma, discussed in the
 appendix. Hence, for $T>T_\text{RW}$,  the asymptotic behavior of the $b_k$ is
  $b_k \sim (-1)^{k-1}/k$, and the prefactor\footnote{To be more precise, the prefactor is the left-sided limit of the density,  $\lim_{\theta_B\nearrow\pi}\text{Im}\chi_1(T,i\,\theta_B)$.} is determined by the
 density at $\theta_B=\pi$.

 The case of a first order transition at some intermediate point, $0<\theta_c<\pi$, on the imaginary
 chemical potential axis can be handled analogously. As mentioned in Section \ref{sec:landau-theory}, this case may correspond to QCD for light or heavy quark masses,
 where the RW endpoint is expected to be a triple point~\cite{forcrand:_const_qcd}. Note,  that in this case the density at $\theta_B=\pi$ vanishes. On either side of the transition, the density is a smooth
 function given  by $f(\theta_B)$ and $g(\theta_B)$, respectively. Thus,
 \begin{align}\label{eq:disc-dens}
  \mathrm{Im}\chi_1^B(T,i\theta_B)& \nonumber\\
  = f(\theta_B)& \Theta(\theta_c-\theta_B)+
  g(\theta_B)\Theta(\theta_B-\theta_c),
 \end{align}
 where $\Theta$ denotes the Heaviside step function.
 After integration by parts, the Fourier coefficients read
 \begin{align}
  b_k&=\frac{2}{\pi}\left(\int_0^{\theta_c} \!\! d\theta_B\,
  f(\theta_B)\sin(k\,\theta_B)\right.\nonumber\\
  &~~~+\left.\int_{\theta_c}^\pi \!\! d\theta_B g(\theta_B)\sin(k\,\theta_B)\right)
  \nonumber \\
  &= \frac{2\cos(k\,\theta_c)}{\pi k}\left( g(\theta_c)-f(\theta_c)
  \right) + \frac{2}{\pi k}\Delta b_k
  \label{eq:bkgeneral}
 \end{align}
where
 \begin{align}\label{eq:deltabk}
  \Delta b_k &=  \int_0^{\theta_c} \!\! d\theta_B f(\theta_B)\cos(k\,\theta_B)\nonumber\\
    &+ \int_{\theta_c}^{\pi} \!\! d\theta_B  g(\theta_B)\cos(k\,\theta_B).
 \end{align}

 For $k\rightarrow \infty$, the term in
 Eq.~\eqref{eq:deltabk} is subleading compared to the first one, owing to the RL
 lemma. Hence, in the case of a first order  transition, the asymptotic form of the Fourier
 coefficients is  $\sim \cos(k\,\theta_c)/k$, with
 amplitude and frequency determined  by the jump in
 density and the location of the transition, respectively.

 At this point, two remarks are called for. First, $b_k \sim k^{-(n+1)}$ is
 expected for a discontinuity in the $n$-th derivative of the
 density~\cite{Tolstov:1976}. The phase of the oscillations depends on whether the
 discontinuity appears in an even or odd derivative. A relevant example is a second-order transition,
 which in the mean-field approximation is associated with a discontinuity in the derivative of the density at
 the critical point.  Thus, such a transition at imaginary $\mu_B$ yields
 high-order Fourier coefficients of the form~\cite{Almasi:2018lok} $b_k \sim \sin(k\, \theta_c)/k^2$.

 Second, we note that the calculation of the Fourier series for the derivative of a discontinuous function must be one done with caution. By differentiating, e.g., the Fourier expansion of the density \eqref{eq:disc-dens} (for real values of the chemical potential), one finds
 \begin{equation}
  \chi_2^B =\frac{\partial \chi_1^B}{\partial \hat{\mu}_B}=\sum_{k=1}^{\infty} c_k\cosh(k\,\hat{\mu}_B),\label{eq:chi2b}
 \end{equation}
 where the Fourier coefficients $c_k=k\, b_k$ are non-zero in the limit\footnote{For the $n$:th derivative of the density ($n>1$), the corresponding Fourier coefficients diverge, $c_k^{(n)}\sim k^{n-1}$.} $k\to\infty$, since $b_k\sim k^{-1}$.  On the other hand, according to the RL lemma, the  Fourier coefficients of a piecewise smooth function vanish asymptotically. This discrepancy is a consequence of the fact that the term-by-term derivative of the Fourier series of an odd function $f(x)$ with period $2\pi$ reproduces that of its derivative $f^\prime (x)$ if and only if the original function is continuous~\cite{Tolstov:1976}.

 \subsection{Second order transition}

 We now consider the effect of a second-order phase
 transition. Here our goal is to extract the leading contribution to the Fourier coefficients $b_k$, induced by the critical singularity.

 In analogy to the mean-field treatment presented in Sect.~\ref{sec:landau}, we parameterize the second-order $O(4)$ critical line near $\mu_B=0$ in terms of the cruvature of the phase boundary $\kappa$,
 \begin{equation}
  T_c(\mu_B)=T_c -\kappa\, \hat{\mu}_B^2.
 \end{equation}
 Using \eqref{eq:singular_f} and the scaling relation $\alpha=2-\beta(\delta+1)$, one finds that, in the chiral limit $(h\to 0)$, the singular parts of the free energy and the baryon density, on the imaginary $\hat{\mu}_B$ axis, are given by
 \begin{align}
  f^{\text{sing}} & \sim -\left| t -\kappa\, \theta_B^2 \right|^{2-\alpha},\,\nonumber \\
  \text{Im}\chi_1^{B,\text{sing}} &\sim ~~\left| t -\kappa\, \theta_B^2 \right|^{1-\alpha} \theta_B,\label{eq:singularpart}
 \end{align}
 where $t$ is the reduced temperature introduced in Sect.~\ref{sec:landau}.
 Thus, at a given temperature $T\geq T_c$, the critical point is located at $\theta_c = \sqrt{t/\kappa}$. The singular contribution to the Fourier coefficients for $T>T_c$ is obtained using \eqref{eq:bk} and \eqref{eq:singularpart},
  \begin{align}
   b_k &\sim A^+\, \int_0^{\theta_c} \! d\theta_B (\theta_c^2-\theta_B^2)^{\phi}\,\theta_B\,\sin\left( k\,  \theta_B  \right) \nonumber \\
&+ A^- \,\int_{\theta_c}^\pi \! d\theta_B (\theta_B^2-\theta_c^2)^{\phi}\,\theta_B\,\sin\left( k\, \theta_B \right)  \label{eq:second-order}\\
   &\sim \Gamma(1+\phi) \left(\theta_c/k\right)^{1+\phi}\left[A^+\,\sin(k\,\theta_c-(1+\phi)\frac{\pi}{2})\right.\nonumber\\
   &+\left. A^- \,\sin(k\,\theta_c+(1+\phi)\frac{\pi}{2})\right],\nonumber
  \end{align}
  where $A^+$ and $A^-$ are the critical amplitudes\footnote{We note that in the mean-field approximation $A^+=0$ and $\phi=1$. Thus, in this case the singular contribution to  the Fourier coefficients is $b_k\sim \sin(k\,\theta_c)/k^2$.} of the density above and below the transition~\cite{Goldenfeld}, respectively and $\phi=1-\alpha$.
  Thus, we obtain a power law, superimposed on an oscillatory term,
  \begin{align}\label{eq:Fourier-2nd-order}
  b_k&\sim A^-\left(\theta_c /k\right)^{2-\alpha}\left[\sin(k\theta_c-\alpha\,\pi/2)\right.\nonumber\\
  &+\left. R_\pm\, \sin(k\theta_c+\alpha\,\pi/2)\right],
  \end{align}
  with the universal ratio $R_\pm=A^+/A^-$. In the $O(4)$ universality class~\cite{Engels:2011km},  $R_\pm\simeq 1.85$ and $\alpha \simeq -0.21$.

  At the $O(4)$ critical point $(T=T_c,\, \theta_c=0)$, the singularity is weaker, the term in (\ref{eq:Fourier-2nd-order}) vanishes and the asymptotic form of the Fourier coefficients is a pure power law
  \begin{equation}
  b_k\sim  k^{2\alpha-4}=k^{-4.42}.
  \end{equation}

  The contribution of the Roberge-Weiss endpoint is evaluated analogously. At intermediate quark masses, this is a
  critical endpoint at $\theta_B=\pi$, belonging to the $Z(2)$
  universality class of the Ising model in three dimensions
  \cite{forcrand:_const_qcd,Bonati:2016pwz}. Hence, the asymptotics of the Fourier coefficients at $T=T_{\text RW}$ is determined by the $Z(2)$ critical exponents.

  Since the RW transition
  line is a symmetry axis for reflections of the phase diagram (cf. Fig.~\ref{fig:phasediagram_theta}) approaching the singularity at constant $T$ corresponds to
  probing the system in the  Ising $h$, i.e., external field,
  direction. Consequently, the singular part of the density  is expected to be characterized by the critical exponent  $\delta$,
  \begin{equation}
   n \sim (\pi-\theta_B)^{1/\delta}.
  \end{equation}
  In this case, the leading singular contribution to the Fourier coefficients is given by
  \begin{align}\label{eq:RW}
   b_k &\sim \int_0^\pi \! d\theta_B\; (\pi-\theta_B)^{1/\delta}\sin(k\,\theta_B) \nonumber \\
   &\sim \frac{(-1)^{k-1}}{k^{1+1/\delta}}.
  \end{align}
  Thus, at $T=T_\text{RW}$, the power law decay of the Fourier coefficients is indeed determined by the critical exponent $\delta$.
  We note that this result is obtained also by setting $\theta_c=\pi$ and $\phi=1/\delta$ in (\ref{eq:second-order}). The fact that the RW endpoint is located at $\theta_B=\pi$, gives rise to the alternating phase of the Fourier coefficients, as for the first-order RW transition in (\ref{eq:RW-1st}). The mean-field case,
  where $\delta=3$, has been discussed in \cite{Almasi:2018lok}.

  For $T < T_c$, the branch point singularities are located on the real $\mu_B$ axis~\cite{Skokov:2010uc}, at $\mu_B=\pm\mu_c$.
  The corresponding singular contribution to the Fourier coefficients $b_k$ is given by
  \begin{align}\label{eq:CEP}
   b_k \sim &\,\, \text{Im}\left[ \int_{0}^{\pi}d\theta_B\,(-t+\kappa\, \theta_B^2)^{\phi}\, \theta_B\,e^{i\,k\,\theta_B} \right]\nonumber\\
   \sim &\,\, \Gamma(1+\phi)\,\left(\hat{\mu}_c/k\right)^{1+\phi}\sin[(1+\phi)\,\pi]\,e^{-k\,\hat{\mu}_c},
  \end{align}
  where again $\phi=1-\alpha$ and $\hat{\mu}_c = \mu_c/T = \sqrt{-t/\kappa}$.
  Thus, the asymptotic contribution of a critical point on the real $\mu_B$ axis is a decreasing
  exponential superimposed on a power law\footnote{In the mean-field approximation ($\alpha=0$) the singular contribution to $b_k$ from a critical point on the real axis vanishes.}. The location of the critical point determines the range of the exponential, while the critical exponent is reflected in the power law. Thus, the corresponding contribution to the asymptotic form of $b_k$, is in general, strongly suppressed.

  The contribution of a possible critical endpoint of the $Z(2)$ universality class~\cite{hatta03:_univer_qcd_critic_and_tricr} and a first-order transition at real values of the baryon chemical potential to the asymptotics of the Fourier coefficients would also be exponentially suppressed.

  We note that for temperatures below the chiral transition at $\mu_B=0$, this is the case also for the part of the Fourier coefficients emanating from  the regular part of the net baryon density. The contribution of the thermal singularities, located at  $\hat{\mu}_B=\pm\hat{m}\pm i\,\pi$, is of the form $b_k  \sim (-1)^{k-1}e^{-k\, m/T}k^{-3/2}$, where $\hat{m}\simeq m_N/T\simeq 6$, since the nucleon is the lightest baryonic degree of freedom in this temperature range. Consequently, for $T<T_c$ it is difficult to extract information on the singularity structure in the complex $\mu_B$ plane from high-order Fourier coefficients.

\subsection{Crossover}

The leading contribution of the crossover singularities~\eqref{eq:mubr_scaling} in the scaling free energy~\eqref{eq:singular_f}, located at $\mu_{br}=T(\pm \hat{\mu}_c\pm i\, \theta_c)$, is obtained by a judicious choice of the integration contour in (\ref{eq:bk}),
\begin{align}\label{eq:crossover}
 b_k
 &\sim e^{-k\hat{\mu}_c}\,\Gamma(1+\phi)\left(\frac{r}{k}\right)^{\phi+1}\nonumber\\
 &\left[A^+\,\sin\left(k\,\theta_c
 -(\phi+1)(\frac{\pi}{2}-\varphi_{br})\right)\right.\\&+A^-\,\left.\sin\left(k\,\theta_c
 +(\phi+1)(\frac{\pi}{2}+\varphi_{br})\right)\right],\nonumber
\end{align}
where $A^+$ and $A^-$ are the amplitudes of the singularity above and below the crossover branch cut and $\phi$ is the corresponding exponent. Moreover, $r=(\hat{\mu}_c^2+\theta_c^2)^{1/2}$ and $\varphi_{br}=\arctan(\hat{\mu}_c/\theta_c)$ is the phase of the crossover branchpoint in the first quadrant.
Consequently, the asymptotic behavior of the Fourier coefficients $b_k$ is that of a damped oscillator superimposed on a power-law dependence on $k$.

For temperatures between the chiral and RW transitions, the imaginary part dominates and the oscillations prevail, while at low temperatures the real part dominates, and the Fourier coefficients are strongly damped with increasing $k$. In the chiral limit, (\ref{eq:crossover}) reduces to (\ref{eq:second-order}) or (\ref{eq:CEP}), for $t>0$ and $t<0$ respectively. The corresponding mean-field behaviour in a QCD-like effective model was discussed in Ref.~\cite{Almasi:2018lok}.

\begingroup
\begin{table*}[!t]
 \caption{Summary of the asymptotic behavior of the Fourier coefficients of the net baryon density,  $b_k$, for different
 classes of the transition. Here $\hat{\mu}_c$ and $\theta_c$ are the real and imaginary parts of the reduced baryon chemical potential at the branch point singularity, $\mu_{\text{br}}/T$. Moreover, $T_{\rm RW}$ is the Roberge-Weiss temperature, $h$ is the strength of the external symmetry-breaking field while $\alpha$ and $\delta$ are critical exponents.}
 \label{tbl:bk}
 \begin{tabular}[b]{l|c|c|c|c|c}\hline
  Temperature range &$T<T_c$&$T=T_c$&$T_c < T <
	      T_{\text{RW}}$&$T=T_{\text{RW}}$&$T>T_{\text{RW}}$
		      \\ \hline
  $h=0$   & $\displaystyle \frac{e^{-k\,\hat{\mu}_c}}{k^{2-\alpha}}$
      \vrule width 0pt height 16pt depth 10pt &
	  $\displaystyle \frac{1}{k^{4-2\alpha}}$ & $ \displaystyle
	      \frac{\sin(k\,\theta_c-\alpha\, \pi/2)+R_\pm \sin(k\,\theta_c+\alpha\, \pi/2) }{k^{2-\alpha}}$ &
		  \raisebox{-5pt}{\multirow{2}{*}{$\displaystyle
		  \frac{(-1)^{k+1}}{k^{1+1/\delta}}$}} &
		  \raisebox{-5pt}{\multirow{2}{*}{$\displaystyle
		      \frac{(-1)^{k+1}}{k}\text{Im}\chi_1^B(\theta_B=\pi)$}}
		      \\ \cline{1-4}
  $h\neq 0$ & \multicolumn{3}{c|}{$\displaystyle
      \frac{e^{-k\,\hat{\mu}_c}(\sin(k\,\theta_c-\alpha\, \pi/2)+R_\pm\sin(k\,\theta_c+\alpha\, \pi/2))}{k^{2-\alpha}}$
      \vrule width 0pt height 16pt depth 10pt  }
      &
		  \\ \hline
 \end{tabular}
\end{table*}
\endgroup

\section{Conclusion}
\label{sec:summary}

In this paper, we have demonstrated that the asymptotic behavior of the
Fourier coefficients of the density at imaginary chemical
potential is, for temperatures above the chiral transition at $\mu_B=0$, governed by the singularities associated with the phase
transition or crossover in the  chemical potential plane.
The asymptotic behavior in the different regimes is summarized in Table
\ref{tbl:bk}. Our results indicate that the singularity structure in the complex chemical potential plane is reflected in the asymptotic behavior of the $b_k$. Indeed, a calculation, within a chiral effective model~\cite{Almasi:2018lok}, demonstrates that in the chiral limit  the location of the critical point, $\theta_c$, can for $T > T_c$ be extracted from a fit to $b_k$ with the expected asymptotic form. Although the exponential damping of $b_k$ leads to
numerical difficulties when the singularity is located far from the imaginary axis, $\mu_c \gg T$, a study of the dependence of the Fourier coefficients on the pion mass may provide information on the  nature of the chiral phase transition, as discussed in Ref.~\cite{Almasi:2018lok}.

In this work, the Fourier decomposition was applied to the net baryon density on
the imaginary $\mu_B$ axis, with applications to QCD matter in mind. Given that
the results presented in this letter are model independent, we expect them to be generic and thus relevant also for other systems with phase transitions or transitions of the crossover type. However, as noted above, the critical contribution to the asymptotics of the Fourier coefficients may be eclipsed by other non-analyticities, located closer to the imaginary $\mu_B$ axis.

In order to apply this approach to LQCD, it would be useful to extend our discussion to systems of finite volume, where the analytic structure in the complex chemical potential plane is different because the cuts and branch points are replaced by Yang-Lee zeros and edge singularities~\cite{Yang:1952be}, respectively. In practice, the calculation of Fourier coefficients at large order in LQCD is certainly numerically very challenging. Nevertheless,
since the location of the singularities depend on the volume as well as on the pion mass, a combined study of the dependence of high-order Fourier coefficients on the volume as well as on the pion mass may provide some insight into the phase structure of QCD matter.

%\acknowledgements
\begin{acknowledgments}
We thank Akira Ohnishi and Vladimir Skokov for fruitful discussions.
The work of AG was supported by the Hungarian OTKA Fund No. K109462 and
by the COST Action CA15213 THOR. K.M. acknowledges support from RIKEN iTHES
project and iTHEM program. K.M. and K.R. acknowledge support by the
 Polish National Science Center NCN,  Maestro grant
DEC-2013/10/A/ST2/00106,  and the Polish "Ministry of Science and Higher Education". This work was supported in part by the Deutsche Forschungsgemeinschaft (DFG) through the grant CRC-TR 211 "Strong-interaction matter under extreme conditions" and by the Extreme Matter Institute EMMI.
K.M and K.R thank Yukawa Institute for Theoretical Physics, Kyoto
 University, where parts of this work were initiated  during ``New
 Frontiers in QCD 2018'' workshop.
\end{acknowledgments}

\appendix*

\section{Large $k$ limit}

In this appendix we derive a relation for the asymptotic form of the Fourier coefficients $b_k$, which is a special case of the Riemann-Lebesgue lemma~\cite{Tolstov:1976}.
For an odd, differentiable function, $f(x)$, the magnitude of the Fourier coefficients for large $k$ decrease as $\sim 1/k$ or faster.
Integrating by parts, one finds
\begin{align}\label{eq:appendix}
 \left|b_k\right| &= \left|\int_0^\pi \; dx f(x) \sin(k\, x) \right|\nonumber \\
 &= \left|\frac{f(0)+(-1)^{k-1} f(\pi)}{k} + \frac{1}{k}\int_0^\pi \; dx f'(x) \cos(k\, x)\right| \nonumber \\
 &\leq \frac{1}{k} \left(\left|f(0)\right|+\left|f(\pi)\right| + \int_0^\pi \; dx \left|f'(x)\right| \right),
\end{align}
where the expression in parantheses is finite, provided the integral of $f'$ is absolutely convergent.
An analogous result is obtained for the Fourier coefficients of even functions.
In Sec.~\ref{sec:bk}, the asymptotic behavior of the Fourier coefficients is determined with the help of the RL lemma.

%\bibliography{HHinteraction,PNJL,PQM,QGPreview,adsqcd,charm,chiral,chpt,eos,exotic,experiment,fluctuation,frg,hydro,immuNotes,inhomogeneous,jpsi_sup,lattice,library,morita,neutronstar,nuclearmatter,pdg,qcd,qgp,qm08,qm12,qm14,qm_conf,statphys,strange,sumrule,test,textbook,tft,thermalmodel}

\end{document}